\documentclass[12pt]{article}
\usepackage{bm}
\usepackage{amsmath}
\usepackage{amssymb}
\usepackage{graphicx} 
\usepackage{wasysym}
\usepackage{hyperref}
\usepackage[top=2cm, bottom=2cm, left=2.5cm, right=2cm]{geometry}

\begin{document}

\title{\bf Mirror magnetic field and its impact on dark matter distribution in galaxies}
\author{Ekaterina Kryukova\thanks{{\bf e-mail}: kryukova.ea15@physics.msu.ru}\\
	\small{\em Department of Physics, Moscow State University}\\
	\small{\em Leninskiye gory 1-2, 119991, Moscow, Russia}}

\maketitle

\begin{abstract}
We obtain the value of the mirror magnetic field during different stages of cosmological evolution. We consider the magnetic field generation in the radiation-dominated era and the post-recombination epoch. We also estimate its galactic low-scale value in the process of dynamo amplification. We discuss a possible effect of the mirror magnetic field on the mirror matter distribution in a galaxy. The model can be generalized by assuming the existence of kinetic mixing between ordinary and mirror particles.
\end{abstract}

\section{Introduction}
\par The cusp-core problem is one of the unsolved problems in modern cosmology. N-body simulations predict density profiles that are sharp in the centre or cusp-like. On the other hand, the observational data show that density profiles of some dwarf galaxies tend to be flat in the centre, hence forming a so-called core (for overview see \cite{block}). It was shown (see \cite{moore}) that the usual CDM model cannot explain this discrepancy. Since then numerous solutions to the cusp-core problem have been proposed: warm dark matter (see ref. \cite{wdm}), self-interacting dark matter (ref. \cite{sidm}), supernovae explosions etc. 
\par One of the possibilities for dark matter origin is the concept of mirror world. Mirror matter particles participate in mirror strong, weak and electromagnetic interactions. Observations show the existence of ordinary galactic magnetic fields. Why should not we suppose the existence of mirror magnetic field then? Usually the impact of electromagnetic field is not considered while modeling matter density distribution in galaxies (see \cite{nelson}, \cite{battaner}, \cite{beck} for attempts). The resulting magnetic field value is connected to matter density. Hence, in mirror world we show the value of mirror magnetic field to be greater than typical galactic value of 1\,$\mu$G for ordinary matter due to higher energy density of dark matter.
\par The purpose of this work is to obtain the value of mirror magnetic field in galaxies and to consider the possible mirror magnetic field impact on the cusp-core problem. Indeed, let us imagine that the initial density distribution in a galaxy is a cusp. Then if we turn on the mirror magnetic field, in the central part of galaxy the mirror charged volume of dark matter will feel the magnetic field pressure that will pull it towards the periphery. Thus the central cusp can become smoother.
\par This paper is organized as follows. In section \ref{overview} we provide an overview of magnetic field generation mechanisms for the ordinary world. Section \ref{mirrorworld} contains the concept of mirror world hypothesis. Generation of mirror magnetic field in RD-epoch is considered in section \ref{harrison}. In section \ref{mishruz} we obtain the value of mirror magnetic field after recombination. Section \ref{dynamo} consists of the estimates of the galactic mirror magnetic field during structure formation. In section \ref{discuss} we discuss probable effect of mirror magnetic field on mirror matter distribution in a galaxy. Finally section \ref{conclude} summarizes the results.
\section{Mechanisms of magnetic field generation} \label{overview}
\par The present value of the large-scale magnetic field in galaxies is about 1 $\mu$G. There are a number of theories describing its possible origin. Here we give a brief overview of them. A large group of theories are so called "battery" mechanisms \cite{zeldovich}. In these theories it is supposed that electrons and ions move with different velocities. The difference in the velocities of charged particles brings about the generation of electric current and hence the growth of magnetic field. The field generation can be based either on the difference in the kinematic viscosities of electrons and ions or on the thermoelectric effect. In general, the value of the magnetic field in these theories is quite small ($10^{-22}-10^{-17}$ G) and could not explain the modern values of $B$ if it were not for achievements of dynamo theory.
\par The dynamo-effect is an effect of self-excitation of magnetic field which occurs when conducting liquid moves in special way. Although there are numerous constraints (for example, the dynamo amplification is impossible in any two-dimensional motion), motions that realize the dynamo effect are quite natural. The dynamo is a process of amplification: it will not work in the absence of a primordial field. The exponential growth of the field in time allows to use battery mechanisms together with the dynamo effect.
\par In this work we base on two battery mechanisms and the dynamo effect to obtain the possible present value of mirror magnetic field. The similar problem for the mirror world is considered in sections \ref{harrison}-\ref{dynamo} along the same lines. 
\par The mechanism proposed by Harrison \cite{harrison} describes the rise of the magnetic field in the radiation-dominated (RD) epoch. An expanding region that possesses angular momentum is considered. The region is uniformly filled with radiation of energy density $\rho_\gamma$ and matter of energy density $\rho$ (ions and non-relativistic electrons) with angular velocities $\omega_\gamma$ and $\omega$ correspondingly. Let $a$ be a scale factor, then in the expanding Universe $\rho a^3=\text{const},~\rho_\gamma a^4=\text{const}$. If there was no interaction between photons, electrons and ions, the angular momentum would be conserved and the expressions $\rho\omega a^5=\text{const},~\rho_\gamma\omega_\gamma a^5=\text{const}$  would hold. Then $\omega\sim a^{-2},~\omega_\gamma\sim a^{-1}$ and we see that in the expanding Universe radiation would slow down less than matter. By considering the interaction of charged particles with photons (Thomson scattering) one obtains that the photon gas entrains electrons to a greater extent than heavy positively charged ions. As a result there is a positively charged ion gas with energy density $\rho$ and an electron-photon gas with energy density $\rho_\gamma$ moving with different angular velocities. This gives rise to an electric current that generates the magnetic field.
\par Mishustin and Ruzmaikin \cite{mishruz} proposed a generation mechanism similar to the previous one. It is supposed to work after recombination in the Dark Ages. The Universe mainly consists of neutral H and He atoms with number density $n_H$ in this period. There are much less free electrons and protons: $n_e=n_p=yn_H$, where $y\simeq3\times10^{-4}-3\times10^{-5}$. After the decoupling of matter from radiation the gas motion can be described relative to a homogeneous radiation background. Let us consider a protogalaxy that expands and rotates relative to the background radiation. Let $\omega(t)$ be the angular velocity of neutral atoms (H and He). The protogalaxy rotation is accompanied by the relative motion of protons and electrons and therefore the magnetic field that arose in the radiation era amplifies. In fact, by taking into account $n_p\ll n_H$ the proton scatterings off neutral atoms of similar masses make the former move with the same angular velocity $\omega(t)$. Radiation nearly does not influence proton motion, but electrons are slowed down by it. The difference in angular velocities of protons and electrons again results in the formation of an electric current and a magnetic field.
\par The main amplification of the galactic magnetic fields to the present value takes place after their formation. The dynamo effect transforms the kinetic energy of conductive medium to the magnetic field energy. At the same time the field value rises with time exponentially. The galactic magnetic field is frozen in ionized interstellar gas as the magnetic Reynolds number $\text{Re}_m$ for gas motion in galaxy at various scales lies within $3\cdot10^5-5\cdot10^7$ \cite{sokolovbook}. In other words, $\text{Re}_m \gg 1$ and gas kinetic energy greatly exceeds the ohmic waste. The dynamo amplification is known to saturate at a certain value. Simplified estimates connect the saturation value with the field equipartition value. At equipartition the equality of kinetic energy of turbulent motion and magnetic field energy is observed. This terminates the magnetic field generation.
\section{Mirror world} \label{mirrorworld}
For the first time some kind of mirror world idea was proposed by Lee and Yang in \cite{leeyang} after parity violation in weak interactions was discovered. Parity symmetry seemed to be natural, and in order to conserve it they assumed the existence of left and right protons that behaved in different ways being produced in $\beta-$decay. 
\par Later this concept was greatly improved and completed. It proposes the existence of mirror particles -- partners of the well-studied Standard Model ones, hence doubling the number of elementary particles. Kobzarev, Okun and Pomeranchuk have shown \cite{okun} that mirror particles cannot take part in the ordinary strong, weak and electromagnetic interactions. At the same time dark matter has to interact with ordinary matter via gravity, otherwise its discovery would be impossible.
\par In the simplest case the mirror symmetry turns out to be exact, and then in the mirror world particles take part in mirror strong, weak and electromagnetic interactions with the same values of coupling constants and charges \cite{foot}. But the mirror world and the ordinary one are not identical. It follows from BBN constraints on number of neutrino flavors $N_\nu$ that the initial temperature of the mirror sector $T'$ has to be lower than the temperature of the ordinary one $T$ (hereinafter quantities related to mirror world are marked by prime). The mixing between the hidden sector and the visible one is assumed to be small, so that one sector does not have any impact except gravitational on dynamics of another one. Since we consider the interaction between the two sectors to be very weak,  temperature ratio $$T'/T=x$$ remains constant at all subsequent epochs of the Universe evolution \cite{oldber}. Generally $x$ is a free parameter, there is only an upper bound on it following from BBN constraints ($\Delta N_\nu < 0.3$, \cite{pdg}): 
\begin{equation}\label{eq:xmax}x<0.47.\end{equation}
If mirror matter gives significant contribution to dark matter density, there is also an upper limit from the formation of large scale structures and CMB spectrum \cite{ber0312}: 
\begin{equation} \label{eq:xmid} x<0.3. \end{equation}
It was also shown in \cite{oldber} that mirror helium abundance is about $75\%$. For detailed description of mirror sector cosmological evolution see \cite{berezhiani}. Modern state of the model and constraints can be found in \cite{newber}. 
\par The mirror protons and electrons are stable, therefore the mirror matter is a natural candidate for the dark matter. In this work it is supposed that mirror matter forms the most part of dark matter. We consider questions related to the mirror electromagnetic interaction, any features of mirror weak and strong interactions are irrelevant in our study. We assume that in the late Universe the mirror matter consist in general of mirror protons $p'$, mirror electrons $e'$ and mirror hydrogen atoms $H'$. Mirror matter interacts with mirror photons $\gamma'$ that together with mirror neutrinos $\nu'$ form mirror radiation. 
\section{Generation of mirror magnetic fields in the RD era}\label{harrison}
The mechanism proposed by Harrison \cite{harrison} gives the magnetic field
\begin{equation}\label{eq:BH}
    \textbf{B}'=-\frac{2m'c}{e}\left(1-\frac{T'}{T_1'}\right)\bm\omega',
\end{equation}
where $m'$ is a coefficient with dimension mass that relates the mirror matter density $\rho'$ and the number of electrons in the unit volume $n_e'$, $T_1'$ and $T'$ are the temperatures at the beginning and at the end of generation, $\omega'$ is the angular velocity of the ion gas in the mirror world. 
\par In the case of the ordinary world dominated by protons and electrons in the RD era, one obtains from the electroneutrality $n_e=n_p$. Then the matter density $\rho=n_e m_e+n_p m_p=n_e (m_e+m_p)\simeq n_e m_H$, so $m=m_H$. In the mirror sector mass fractions of mirror helium $He'$ and mirror hydrogen $H'$ are $75\%$ and $25\%$ respectively. The condition of electroneutrality takes the form $n_e'=n_p'+2n_{He^{++}}'$, the expression for matter density is $\rho'=n_e' m_e+n_p' m_p+n_{He^{++}}'m_{He^{++}}$. One can solve the system of equations and obtain $m'=\rho'/n_e'$:
\begin{equation} \label{eq:m} m'=\frac{m_{He^{++}}m_H+3m_p m_{He}}{m_{He^{++}}+6m_p}\simeq \frac{4m_{He}m_H}{m_{He}+6m_H}.  \end{equation}
Finally after substituting numerical values we get $m'=1.6m_H$. As the value of $m'$ is of the same order in the both cases of hydrogen and helium prevalence, hereinafter we will not consider the presence of helium.   
\par Generation starts when mirror electrons become non-relativistic:  
\begin{equation} \label{eq:T1}
    T_1'=m_ec^2/k=6\times10^9K,
\end{equation} and finishes when energy densities of mirror matter (that is dark matter by assumption) and mirror photons are equal (equality in the mirror world): 
\begin{equation}\label{eq:equality}
    \Omega_d\left(\frac{a_0}{a}\right)^3=\Omega_\gamma'\left(\frac{a_0}{a}\right)^4,
\end{equation}
where $\Omega_d$, $\Omega_\gamma'$ are present density parameters of dark matter and mirror photons respectively.
\par Let us evaluate the energy density of mirror photons  $\rho_\gamma'$. The ratio between the temperatures of our and mirror world conserves \cite{berezhiani}: $x=T'/T=T_0'/T_0=\text{const}.$ Since the photon energy density is proportional to $T^4$, in every moment of time \begin{equation}\label{eq:rho}
    \frac{\rho_\gamma'}{\rho_\gamma}=x^4,\hspace{20pt} \frac{\Omega_\gamma'}{\Omega_\gamma}=x^4.
\end{equation}
Then from the equality of energy densities \eqref{eq:equality} we find that the generation finishes at: \begin{equation} \label{eq:a}
    \frac{a_0}{a}=\frac{\Omega_dh^2}{\Omega_\gamma'h^2}=4760x^{-4}=\frac{T'}{T_0'}.
\end{equation} Using $T_0=2.73K$, we obtain the mirror world temperature at that time: \begin{equation}\label{eq:T}
    T'=13\times10^3 x^{-3} K.
\end{equation}
An inherent constraint on $x$ in our model is that the initial temperature \eqref{eq:T1} must be higher than the final one \eqref{eq:T}: $T_1'\geq T'$, that is \begin{equation}
    x>0.013.
\end{equation}
The radiation energy density (photons and neutrinos) is connected to temperature as $\rho_{rad}={\pi^2}/{30}\,g_*T^4,$ where $g_*$ is the effective number of degrees of freedom \cite{rubakov} that does not change in both mirror and ordinary worlds in the considered period of time. We estimate $\omega'$ similar to \cite{harrison}. For the mirror world at radiation dominated era $\omega'\simeq\omega_\gamma'$, so
\begin{equation} \label{eq:w}
    \omega'\simeq\sqrt{\frac{8\pi G(\rho_{rad}+\rho_{rad}')}{3}}=\sqrt{\frac{8\pi G\rho_{rad}(1+x^4)}{3}}.
\end{equation} 
The modern energy density of photons is $\rho_{\gamma0}=2.55\times10^{-10}$GeV/cm$^3$ \cite{rubakov}. Then at the end of generation according to formula \eqref{eq:a}, 
\begin{equation}\label{eq:ro}
    \rho_{rad}=1.68\rho_{\gamma0}\left(\frac{a_0}{a}\right)^4=3.9\times10^{-19}x^{-16}\text{g/cm}^3.
\end{equation}
From formulae \eqref{eq:BH}, \eqref{eq:T1}, \eqref{eq:T}, \eqref{eq:w}, \eqref{eq:ro} we obtain:
\begin{equation}\label{eq:firstB}
    B'=-9.4\times10^{-17}x^{-8}\left(1-2.2\times10^{-6}x^{-3}\right)\text{G.}
\end{equation}
\begin{figure}
	\centering
	\includegraphics[width=0.7\linewidth]{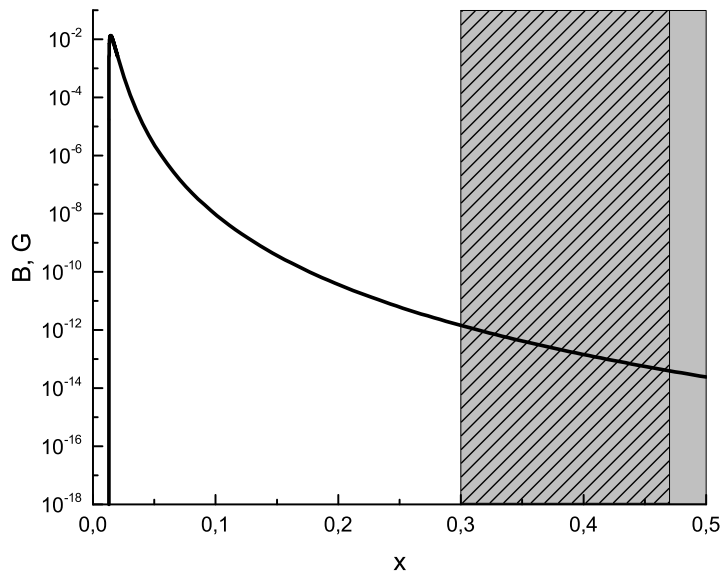}
	\caption{The magnetic field obtained by Harrison mechanism in the end of RD epoch versus parameter $x$\label{graph1}.}
\end{figure}
\par The dependence of the mirror magnetic field $B'$ on the parameter $x$ is shown in figure \ref{graph1}. The area where $x>0.47$ \eqref{eq:xmax}, that is forbidden in this model, is darkened. We also mark with the oblique line the area where $0.3<x<0.47$ \eqref{eq:xmid}, that is forbidden in the case if mirror matter forms the dominant part of dark matter. The contribution of correction  $\left(1-\frac{T'}{T_1'}\right)$ in \eqref{eq:firstB} slightly differs from 1 when $x\gg0.014$ ($x=0.014$ corresponds to the maximum point in figure \ref{graph1}), therefore hereinafter we will neglect it.
\par We should notice that the value of mirror magnetic field obtained within the bounds of Harrison mechanism is greater than one in the ordinary world since mirror world temperature is lower ($x<1$).
\par The idea of Harrison mechanism proposed in 1970 was subsequently used in several mechanisms of the magnetic field generation in the radiation-dominated era (for instance, see reviews \cite{rev1712}, \cite{rev0312}). We should notice that there exists an argument against the original Harrison mechanism. The thing is that the assumption of primordial vorticity requires the existence of the first order vector perturbations in this epoch \cite{rev1303}. They are not generated if the matter perturbations are originated solely from inflation, but even if they are, the vorticity in the RD era would decay rapidly with the expansion of the Universe \cite{rev0207}. For the criticism of Harrison mechanism also see the ref. \cite{rees}. 
\par As it was already stated, there are several equivalent mechanisms following the idea of Harrison mechanism in the RD era. All of them exploit the Thomson scattering to realize the interaction between photons and electrons. As the physical processes stay the same and we have the power dependence of magnetic field on temperature, it is unlikely that the change of  model inside the class of works based on Harrison mechanism will result in the drastic change in the seed of magnetic field generated at the radiation-dominated epoch.

\section{Mirror magnetic field after recombination}\label{mishruz}
\par For the mirror world the Friedmann equation takes the form:
\begin{equation}\label{eq:frid}
    \left(\frac{\dot a}{a}\right)^2=\frac{8\pi G}{3}\rho_c\left((\Omega_b+\Omega_d)\left(\frac{a_0}{a}\right)^3+(\Omega_{rad}+\Omega_{rad}')\left(\frac{a_0}{a}\right)^4+\Omega_\Lambda\right),
\end{equation}
where $a$ is a scale factor, $a_0$ -- its present value, $\rho_c$ -- critical energy density, $\Omega_b$, $\Omega_{rad}$ ($\Omega_d$, $\Omega_{rad}'$) -- present density parameters of ordinary (mirror) matter and radiation respectively \cite{rubakov}. 
\par We begin from estimating important instants for the second (Mishustin and Ruzmaikin) mechanism. This mechanism starts working after photon decoupling. The main assumptions are broken when Hubble expanding changes to self compression by gravity. A prospective galaxy or galaxy cluster starts to form in the considered region. To estimate the value of \textbf{B}$'$ we need to know the moment of matter decoupling from radiation ($t_r$ and $a_r$), the initial moment of structure formation ($t_*$ and $a_*$), $t(a)$ and $\omega(a)$ dependence. \par At the moment of matter-radiation equality energy densities of all (ordinary and mirror) matter and all radiation are equal \cite{berezhiani}.
$$\Omega_m=\Omega_b+\Omega_d\hspace{25pt}\Omega_{r}=\Omega_{rad}+\Omega_{rad}'$$
$$\Omega_m\left(\frac{a_0}{a}\right)^3=\Omega_r\left(\frac{a_0}{a}\right)^4.$$
Using relation \eqref{eq:rho}, expression for present radiation density parameter \cite{rubakov}: $\Omega_{rad}h^2=1.68 \Omega_\gamma h^2=4.2\times10^{-5},$ and Planck experiment data \cite{planck}: $\Omega_b h^2=0.022, \Omega_d h^2=0.119,$
we get
\begin{equation}\label{eq:eq}
    a_{eq}=a_0\frac{\Omega_m}{\Omega_r}=a_0\frac{1+x^4}{3360}.
\end{equation}
\par Let us estimate the moment when mirror photons decouple from matter (last scattering). In the ordinary Universe the decoupling takes place at the temperature $T_{dec}\simeq0.26\text{ eV}$, which corresponds to the redshift $1+z_{dec}=T_{dec}/T_0\simeq1100.$ In \cite{berezhiani} it is shown, that the temperature of decoupling for the mirror sector can be evaluated in the same way as for the ordinary one, that is why $T_{dec}'\simeq T_{dec}$, and the redshift equals
\begin{equation*}
    1+z_{dec}'=\frac{a_0}{a_{dec}'}=\frac{T_{dec}'}{T_0'}\simeq\frac{T_{dec}}{xT_0}=x^{-1}(1+z_{dec})\simeq1100x^{-1}.
\end{equation*}
The scale factor at the decoupling
\begin{equation}\label{eq:r}
    a_r=\frac{a_0x}{1100}.
\end{equation}
\par From expressions \eqref{eq:eq} and \eqref{eq:r} one can see that moments of photon decoupling and matter-radiation equality and their order depend on parameter $x$. Recombination in the mirror world occurs before matter-radiation equality if $a_r<a_{eq}$, which implies for \begin{equation}\label{eq:case} x<0.33. \end{equation}
Thus, at $x<0.33$ the mechanism goes consequently through RD- and MD- epochs, and at $x>0.33$ the field growth occurs only in MD-epoch. This is crucial for integrating in \eqref{eq:BMR} -- one will have to choose different regimes $t(a)$ for stages of radiation and matter dominating.
\par We rewrite the expression for magnetic field from paper \cite{mishruz} in terms of mirror world: \begin{equation}\label{eq:BMR}
    \textbf{B}'(t)=\frac{1}{\beta a^2(t_*)}\left(\int\limits_{t_r}^{t_*}\frac{2a^2\bm\omega'}{\tau_{e\gamma}'}\,dt+(a^2\beta\textbf{B}')|_{t=t_r}\right).
\end{equation} 
Here $a$ is a scale factor, $\tau'_{e\gamma}$ is a characteristic time of Thomson scattering, $\sigma_T$ is its cross-section, $\sigma_T=6.65\cdot10^{-25}\text{cm}^2$, \begin{equation} \label{eq:tau}
    \beta=\frac{e}{m_ec},\hspace{20pt} \frac{1}{\tau'_{e\gamma}}=\frac{4\sigma_T\rho'_\gamma c}{3m_e}.
\end{equation} 
The upper integration limit in equation \eqref{eq:BMR} is given by the start of structure formation. We define $t_*$ as an initial moment of formation of galaxy with mass $M$ in order to keep the kind of $t(a)$ dependence unchanged at the whole integration interval in MD-epoch. Let $R$ be the present size of spatial region with density $\rho_{m0}$ that contains matter of mass $M$. We determine $R$ from  $M={4\pi}/{3}\,R^3\rho_{m0},$ where $\rho_{m0}=\Omega_m \rho_c=3.6\times10^{10}M_{\astrosun}/\text{Mpc}^3.$ The size $R=1$ Mpc corresponds to medium-sized galaxy of mass $M=1.5\times10^{11}M_{\astrosun}$, and $R=400$ kpc corresponds to mass of the largest dwarf galaxies $M=1.0\times10^{10}M_{\astrosun}.$ Now we estimate $a_*$.
\par According to book \cite{rubakov2}, we estimate the variance of smoothed density contrast
\begin{equation} \label{eq:int}
    \sigma_R^2(t)=\int \limits_{0}^{\infty}\frac{dk}{k}\mathcal{P}(k, t)\frac{9j_1^2(kR/a_0)}{(kR/a_0)^2}.
\end{equation}
Here we omit the calculations and give the approximate formula that was used in this work:
\begin{equation}
    \sigma_{R}^2(z)\simeq \frac{1}{(1+z)^2}\left(0.57\ln^2\left(\frac{26\,\text{Mpc}}{R}\times1.57\right)+\frac{1}{3}\ln^3\left(\frac{26\, \text{Mpc}}{R}\right)+1.4\right).
\end{equation}
Using it for $\sigma_R=0.5$, that corresponds to moment when only a small percentage of all galaxies was gravitationally bound, and for $R$ obtained earlier we get for a dwarf galaxy 
\begin{equation} \label{eq:z1}
    R=400\text{ kpc}, \hspace{0.5cm}z_{*}=11.3,
\end{equation} for medium-sized galaxy 
\begin{equation} \label{eq:z2}
    R=1\text{ Mpc}, \hspace{0.5cm}z_{*}=8.1.
\end{equation}
\par According to \cite{galform}, matter density of a galaxy $\rho_g'$ at the moment of virialization  relates with the surrounding matter density as
\begin{equation}
    \rho_g'\simeq150\rho_{m*}=150\rho_c\Omega_m\left(\frac{a_0}{a_*}\right)^3.
\end{equation}
\par Let us estimate the angular velocity of mirror protogalaxy $\omega_*'$. We introduce $\omega_{g0}'$, $R_{g0}$ -- present galaxy angular velocity and halo radius, $R_{g*}$ -- protogalaxy halo radius. Then we use the conservation of angular momentum for the process of protogalaxy compression and the conservation of total matter mass:
\begin{equation}
    \omega_*'= \omega_{g0}'R_{g0}^2\left(\frac{200\pi\Omega_m\rho_{c}}{M}\right)^{2/3}\left(\frac{a_0}{a_*}\right)^2.
\end{equation}
\par Finally we express time of the protogalaxy formation $t_*$. For that we solve the Friedmann equation  \eqref{eq:frid} for typical values of $a_*$. Formulae \eqref{eq:z1}, \eqref{eq:z2} give us $z_{*}=8-11$. Thus for $a_*$ being considered it is useful to take into account the impact of matter and the cosmological $\Lambda$-constant. $$\dot a^2=H_0^2\left(\frac{\Omega_m a_0^3}{a}+\Omega_{\Lambda}a^2\right).$$
We get the expression for time of structure formation
\begin{equation}
    t_*=\frac{2}{3\sqrt{\Omega_{\Lambda}}H_0}\text{Arsh}\sqrt{\frac{\Omega_{\Lambda}}{\Omega_m}\left(\frac{a_*}{a_0}\right)^3}.
\end{equation}
For values of $x$ determined by expression \eqref{eq:case} the amplification takes place both at the radiation era and at the matter dominated epoch. We rewrite \eqref{eq:BMR} using introduced variables.
\begin{equation}\label{eq:BMMR}
    \textbf{B}'(t)=\frac{1}{\beta a_*^2}\left(\int\limits_{t_r}^{t_{eq}}+\int\limits_{t_{eq}}^{t_*}\right)\frac{2a^2\bm\omega'}{\tau_{e\gamma}'}\,dt+\left(\frac{a_I}{a_*}\right)^2\textbf{B}_I=\textbf{B}_1+\textbf{B}_2+\left(\frac{a_I}{a_*}\right)^2\textbf{B}_I.
\end{equation}
Here $a_I$ is the scale factor at the end of work of Harrison mechanism, $B_I$ is the corresponding magnetic field. We use formulae \eqref{eq:a} and \eqref{eq:firstB} assuming that $x\gg0.014$,
\begin{equation}
    a_I=\frac{a_0x^4}{4760}, \hspace{30pt} B_I=-9.4\times10^{-17}x^{-8}\text{G}.
\end{equation}
Using expressions \eqref{eq:tau}, \eqref{eq:rho} and the angular momentum conservation, we obtain: \begin{equation}
    \frac{2a^2\omega'}{\tau_{e\gamma}'}=\frac{8\sigma_Tc}{3m_e}\rho_{\gamma0}x^4\frac{a_0^4}{a^4}a_*^2\omega_*'.
\end{equation}
For $t_{r}<t<t_{eq}$: $$t=a^2\cdot \text{const}\hspace{30pt}dt=t_{eq}\frac{2ada}{a_{eq}^2},$$
for $t_{eq}<t<t_*$: $$t=a^{3/2}\cdot \text{const}\hspace{30 pt}dt=t_*\frac{3a^{1/2}da}{2a_*^{3/2}}.$$
Let us introduce the factor $F$ to simplify the expressions for $B_i$:
$$F=\frac{8\sigma_Tc^2}{e}\rho_{\gamma0}x^4\omega_*'t_*\left(\frac{a_0}{a_*}\right)^{3/2}.$$
After integrating in \eqref{eq:BMMR} we obtain:
\begin{equation}\label{eq:B1}
    B_1=\frac{F}{3}\left[\left(\frac{a_0}{a_r}\right)^{2}\left(\frac{a_0}{a_{eq}}\right)^{1/2}-\left(\frac{a_0}{a_{eq}}\right)^{5/2}\right],
\end{equation}
\begin{equation}\label{eq:B2}
    B_2 = \frac{F}{5}\left[\left(\frac{a_0}{a_{eq}}\right)^{5/2}-\left(\frac{a_0}{a_*}\right)^{5/2}\right].
\end{equation}
Similarly to expression \eqref{eq:BMMR} for $x>0.33$, 
\begin{equation}\label{eq:BMMMR}
    \textbf{B}'(t)=\frac{1}{\beta a_*^2}\int\limits_{t_{r}}^{t_*}\frac{2a^2\bm\omega'}{\tau_{e\gamma}'}\,dt+\left(\frac{a_I}{a_*}\right)^2\textbf{B}_I=\textbf{B}_3+\left(\frac{a_I}{a_*}\right)^2\textbf{B}_I.
\end{equation}
By repeating calculations in the previous paragraph (compare with \eqref{eq:B2}), we find
\begin{equation}
    B_3=\frac{F}{5}\left[\left(\frac{a_0}{a_{r}}\right)^{5/2}-\left(\frac{a_0}{a_*}\right)^{5/2}\right].
\end{equation}
\par As examples, for M33 galaxy by substituting $v'=120\,\text{km}\cdot\text{s}^{-1}$, $ M'=6\times10^{10}M_{\astrosun}$, $R_{g0}=17\,\text{kpc}$ \cite{M33}, we obtain
$a_0/a_*=10.1$, $\omega_*'=3.5\times10^{-16}\,\text{s}^{-1}$, $t_*=1.7\times10^{16}\,\text{s}$.
\par
\begin{figure}
	\centering
	\includegraphics[width=0.7\linewidth]{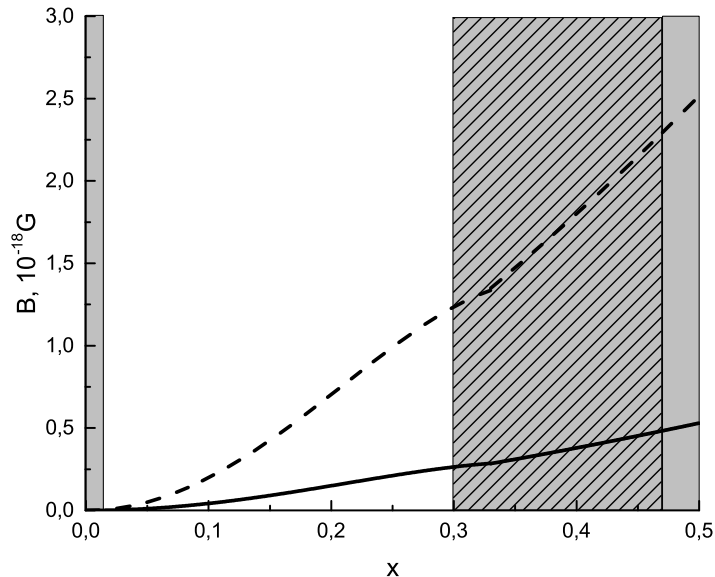}
	\caption{The magnetic field obtained by Mishustin and Ruzmaikin mechanism in the post-recombination epoch versus parameter $x$\label{graph2} for M33 galaxy (dashed line) and M31 galaxy (solid line).}
\end{figure}

 For M31 galaxy $v'=190\,\text{km}\cdot\text{s}^{-1}, M'=1.2\times10^{12}M_{\astrosun}$, $R_{g0}=35\,\text{kpc}$ \cite{M31} and we get the following values: \(a_0/a_*=7.0,\,\omega_*'=7.5\times10^{-16}\,\text{s}^{-1},
t_*=3.0\times10^{16}\,\text{s}.\)
The dependence of the mirror magnetic field on parameter $x$ for M33 and M31 galaxies at the moment of the inhomogeneity separation is shown in figure \ref{graph2}. The area that is forbidden in this model is darkened and marked by oblique line. Curves consist of two parts: for $x<0.33$ and $x>0.33$. 

\section{Dynamo amplification of mirror magnetic field}\label{dynamo}
In the theory of the dynamo effect magnetic field lines are expected to be frozen into the interstellar medium. As it was mentioned above, the condition $\text{Re}_m\gg1$ can be treated as a criterion of the magnetic flux conservation and thus a criterion of magnetic field lines freezing-in. Here $\text{Re}_m$ is the magnetic Reynolds number defined as
\begin{equation} \label{eq:re}
    \text{Re}_m=\frac{ul\cdot4\pi\sigma}{c^2},
\end{equation}
where $u$ is the typical turbulent velocity and $l$ is the length scale of turbulence of conducting liquid, $\sigma$ is its conductivity \cite{land8}. For the mirror world from general considerations we can assume $u'\sim u,\,l'\sim l.$ The turbulent velocity of the interstellar gas is estimated by the gas velocity dispersion in galaxies. For galaxies with various rate of star formation this value remains approximately constant and equal to $u=10$ km/s \cite{ism}. The basic length scale of turbulence in galaxies is taken to be $l=100$ pc. For the dark matter halo in \cite{sokolovart} the following values are taken: $u'=20 \text{ km/s}$ and $l'=200$ pc. 
\par Since the temperature of the mirror world $T'$ is lower than the temperature of ordinary one $T$, the possibility of star formation and reionization stays unclear (for more detailed discussion see \cite{blinnikov}). We assume that these processes proceed with lower intensity, thus the interstellar medium is filled with the partly ionized hydrogen plasma, for which the degree of ionization and the temperature of electron component are free parameters. 
\par Let us estimate the conductivity of the partly ionized plasma and conclude on the relevance of dynamo theory for the mirror world. Let $n_{e}', n_{p}', n_{H}', v_{e}', v_{p}', v_{H}'$ be the concentrations and the velocities of the corresponding mirror particles. Then the current density $j$ is defined in a standard way:
\begin{equation}\label{eq:j}
    j=(-n_{e}'v_{e}'+n_{p}'v_{p}')e.
\end{equation}
The interaction with protons is crucial for electron motion, the interaction with electrons and elastic collisions with hydrogen atoms are important for proton motion. We write the equations of motion of electrons and ions assuming that the Lorentz force compensates the slowing down via momentum exchange with other particles (the magnetic component of Lorentz force is neglected, taking it into account in \cite{land10} gives the value of conductivity of the same order):
\begin{equation}\label{eq:ee}
    -eE=m_ev_{e}'\nu_{ep}'\hspace{1.5 cm}eE=m_pv_{p}'(\nu_{pe}'+\nu_{pH}'),
\end{equation}
where $\nu_i'$ are the typical frequencies of momentum transfer to the corresponding particles. 
By substituting equations \eqref{eq:j} and \eqref{eq:ee} into the Ohms law $j = \sigma E$ we get the expression for $\sigma$:
\begin{equation}
    \sigma=e^2\left[\frac{n_e'}{m_e\nu_{ep}'}+\frac{n_p'}{m_p(\nu_{pe}'+\nu_{pH}')}\right].
\end{equation}
\par The frequency of proton-electron collisions is given by formula \cite{land10} $$\nu_{ep}'=\frac{4\pi e^4 n_e' L_e}{(T'_{e})^{3/2}m_{e}^{1/2}},$$ where $L_e$ is the Coulomb logarithm, $L_e\sim10$, $T_e'$ is the temperature of the electron component in energy units, the total cross-section for elastic collisions of protons and hydrogen atoms $\sigma_{pH}\sim10^{-15}\text{ cm}^2$, so that $$\nu_{pH}'=\sigma_{pH}n_H'v_p'.$$
Given that we estimate the conductivity in explicit way:
$$\sigma=\frac{(T_e')^{3/2}}{4\pi e^2 L_e m_e^{1/2}}\left[1+\frac{m_e}{m_p}\frac{1}{\left(1+\sigma_{pH}\frac{n_H'}{n_p'}\frac{(T'_e)^2}{4 \pi e^4 L_e}\right)}\right].$$
\par Let us find the degree of ionization at which the right item in round brackets is not more than 0.01 and, hence, makes a small contribution to the plasma conductivity.
\begin{equation} \label{eq:ineq}
    \frac{n_p'}{n_H'}\geq\frac{100\sigma_{pH}(T_e')^2}{4\pi e^4L_e}=0.038(T'_e/\text{eV})^2.
\end{equation}
For degrees of ionization at which the inequality \eqref{eq:ineq} holds, we can take 
\begin{equation} \label{eq:sigmacond}
    \sigma=\frac{(T_e')^{3/2}}{4\pi e^2 L_e m_e^{1/2}}.
\end{equation}
Using the expression for the conductivity \eqref{eq:sigmacond} and the definition of the magnetic Reynolds number \eqref{eq:re}, it is possible to express the temperature of electron component needed for the dynamo amplification: $\text{for } Re_{m}\geq100$ it has to be $T'_e\geq6\times10^{-8}$K, this certainly takes place in the mirror world. So the exponential amplification of the mirror magnetic field with time is possible.
\par The field growth starts from the dynamo amplification at low scales within one turbulence cell (the so-called fast dynamo proposed by Zel'dovich). According to the equipartition condition $\rho u^2/2=B^2/8\pi$ the saturation is reached at low-scaled field \begin{equation} \label{smallscale}
    b'=u'\sqrt{4\pi \rho_d}=5.0\times10^{-5}\text{ G},
\end{equation} where $u'=20$ km/s is a typical turbulent velocity for the dark matter, $\rho_d$ is the dark matter density in galaxy, $\rho_d\simeq5\rho_b=5\times10^{-23}\text{ g/cm}^3$  \cite{sokolovart}. The field growth according to law $B'=B'_0e^{t/\tau_{s}}$ is determined by e-folding time $\tau_{s}=l'/u'=3.1\times10^{14}$ s. Then the growth of the magnetic field from the value $B'_0\simeq10^{-18}$ G (see figure \ref{graph2}) to saturation value $b'$, described by formula \eqref{smallscale}, takes $t_1=\tau_s\ln{({b'}/{B'_0})}=3.1\times10^{8}\text{ years}.$
\par Large-scale field, determined by averaging small-scale fields over all $N$ turbulent cells, is given by formula $B'=b'N^{-1/2}$ \cite{sokolovart}. The amplification of this field is possible up to the maximum value $B_{eq}'$ that might depend on the halo size, the orbital speed of mirror matter in galaxy and the mirror matter density. The further research requires numerical modelling. Some rough estimates performed for ordinary matter (see \cite{sokolovart}) show that $B_{eq}'\sim b'$, they are of the same order of magnitude and that $B_{eq}'$ appears long before the present time. 
\section{Discussion}\label{discuss}
Now we qualitatively discuss the possible effect on the dark matter density profile. As stated by formula \eqref{smallscale} the maximum value of the mirror magnetic field is defined by the dark matter density. Let the mass distribution be typical for galaxies with a cusp in the center. One can take the Navarro-Frenk-White profile \(\rho(r)=\rho_{0}/[{r/R_{s}}\left(1+{r/R_{s}}\right)^{2}]\). We consider the simplified case at which the mirror magnetic field is defined by the given density profile. So we neglect the influence of the magnetic field on the initial mass distribution in galaxy. 
\par Thus as $B\sim\sqrt{\rho_d}$, the value of the mirror magnetic field and hence of the magnetic field energy will be bigger in the central region rather than at large radii. The presence of additional energy density will cause the pressure on the given probe charged volume moving in the equilibrium gravitational potential. Then there will be the flux of particles moving from centre to periphery and thus the central cusp will become smoother.
\par It can be easily checked that this effect does not cause a total mass loss of the galaxy as at the large distances from the galactic centre the mirror magnetic field influence on the particle motion is much less than the influence of the gravitational field. One can compare the energy densities of the electro-magnetic field: $B^2/8\pi\sim\rho_du'^2/2$ and the kinetic energy density: $\rho_d v'^2_{orb}/2$, where $u'=20$ km/s is the typical velocity of the turbulent motion in galaxies and $v'_{orb}$ is the mirror matter orbital velocity. Here we suppose that orbital velocities are determined mostly by the gravitational field and thus the kinetic energy of a unit volume represents the impact of the gravitational field. At the border where orbital velocities can reach values of 100 km/s the kinetic energy would exceed the magnetic one in about 25 times. This is a possible reason to prevent the dark matter from escaping the galaxy. 
\par It is clear that the mirror magnetic field becomes significant at small radii where $u'\sim v'_{orb}$. According to rotation curves for M31 and M33 this happens at radii less than 5 kpc. The effect of central density smoothing can be useful in considering the core-cusp problem.
\par The considered model can be generalized basing on the idea of magnetic field in dark sector. At late stages of the Universe evolution the existence of dark matter with vector fields is required. In order that the mirror electromagnetic interaction be long-range, this field should be massless. Within the bounds of this approach there should be the mirror magnetic field, but the charged particles masses do not have to coincide with their values in ordinary world. The further generalization can include the kinetic mixing between ordinary and mirror photons. It should be investigated extra if this mixing can be used to generate the mirror magnetic field.  
\section{Results}\label{conclude}
A possible scenario of generation and amplification of galactic mirror magnetic field was considered. The obtained results given that $x=1$ agree with the predictions of magnetic field value for the ordinary world. The maximum value of mirror magnetic field for medium-sized M33 galaxy turns out to be about 50 times greater than typical ordinary galactic magnetic field of $1\mu G$. This allows us to discuss its influence on the mirror matter distribution in galaxies. On qualitative level we derive that it can make the central cusp smoother. This effect outlines another approach to the core-cusp problem.

The author is grateful to D.~S.~Gorbunov (INR RAS) for suggesting the problem, helpful discussions and valuable comments on this work.

\end{document}